\documentclass[article]{revtex4}
\usepackage[latin1]{inputenc}
\usepackage{amsmath}
\usepackage{amssymb}
\usepackage{indentfirst}
\usepackage{epsfig}
\usepackage{amsthm,amscd}
\usepackage{bm}
\usepackage{url}            
\usepackage{fleqn}
\usepackage{natbib}

\newcommand{\rme}{\mathrm{e\,}}
\newcommand{\rmi}{\mathrm{i\,}}

\def\p{\partial}

\def\erf{{\rm erf}\,}
\def\exp{{\rm exp}\,}

\newcommand{\Sh}{Schr\"odinger\ }

\catcode`\@=11
\renewcommand\footnoterule{%
  \kern-3\p@
  \hrule\@width.4\columnwidth
  \kern2.6\p@}
\renewcommand\@makefntext[1]{%
    \parindent 1em\noindent
    \hb@xt@1.8em{\hss$^{\@thefnmark}$)}\hspace{2pt}%
    \footnotesize\rmfamily#1}  
\def\@makefnmark{\hspace{.5pt}\hbox{$^{\@thefnmark}$%
\hspace{-1pt})}} 

\begin{document}

\title[Isochronous an-harmonic oscillators, propagators and x-Mehler formula]
{Propagators of isochronous an-harmonic oscillators and Mehler formula for the x-Hermite polynomials}

\author{Andrey M. Pupasov-Maksimov}
 \email{pupasov.maksimov@ufjf.edu.br}
\affiliation{%
Depto. de Matem\'atica, ICE,\\
Universidade Federal de Juiz de Fora, MG, Brasil
}%

\date{\today}

\begin{abstract}
It is shown that 
fundamental solutions $K^\sigma(x,y;t)=\langle x|\rme^{-\rmi H^\sigma t}|y\rangle$ of the non-stationary Schr\"{o}dinger 
equation (Green functions, or propagators)
for the rational extensions 
of the Harmonic oscillator $H^\sigma=H_{\rm osc}+\Delta V^\sigma$ 
are expressed in terms of elementary functions only. An algorithm to 
calculate explicitly $K_\sigma$ for an arbitrary $\sigma\in \mathbb{N}$
is given, compact expressions for $K^{\{1,2\}}$ and $K^{\{2,3\}}$ are presented. 
A generalization of the Mehler's formula to the case of exceptional Hermite polynomials 
is given.
\end{abstract}

\pacs{Valid PACS appear here}
\maketitle

\section*{Introduction \label{sec:intro}}


In this work we present new examples of 
exactly-solved propagators in one-dimensional 
quantum mechanics.  
We consider rationally extended Harmonic oscillators \cite{gomez2014rational}. 
In this case the evolution of wave packets is periodic, we have so called 
isochronous anharmonic oscillators \cite{carinena2007isochronous}. This periodicity is related with 
the quasi-equidistant structure of the spectrum of the Hamiltonian \cite{dubov1994equidistant}. 

The simplest rational extension is given by  the potential \cite{bagrov1997darboux} 
\begin{equation}\label{def:potential-V12-dimensionless}
V^{\{1,2\}}[x]=\frac{x^2}{4}+2\left(1 + 2\frac{(x^2-1)}{(x^2+1)^2}\right) \,,
\end{equation}
which leads to the quasi-equidistant spectrum \cite{dubov1994equidistant}, $E_n=n+\frac{1}{2}$, where $n\in \mathbb{N}_0\setminus \{1,2\}$, for the stationary Schr\"{o}dinger 
equation.

Another example is the two-well perturbation of the oscillator 
\begin{equation}\label{def:potential-V23-dimensionless}
V^{\{2,3\}}[x]=\frac{x^2}{4}+2\left(1 +4x^2\frac{x^4-
9}{(x^4+3)^2}\right) 
\end{equation}
with the  quasi-equidistant spectrum, $E_n=n+\frac{1}{2}$, $n\in \mathbb{N}_0\setminus \{2,3\}$.

Note that each rational extension $V^\sigma$ of the Harmionic oscillator is defined by a sequence of levels $\sigma\in \mathbb{N}$ which are deleted from the spectrum by Darboux-Crum transformations \cite{gomez2014rational}.
Darboux transformations represent a powerful tool to manipulate physical properties of one-dimensional 
quantum systems,   \cite{andrianov2012nonlinear}.  
In the case of shape-invariant potentials, relations between propagators were studied in \cite{das1990propagators,PhysRevD.47.4796}. 
A more general approach for the calculations of propagators 
for potentials generated by an arbitrary chain of Darboux transformations
were developed in \cite{pupasov2005exact,samsonov2005susy,samsonov2006exact,pupasov2007exact}. 

As a particular example,
propagators $K^{\{k,k+1\}}$ for the $V^{\{k,k+1\}}$ family
were defined through a generating function $S(x,y;t|J)$ which contains the error-function \cite{pupasov2007exact}. 
Here we will further simplify and extend this result. First, we will show that 
the propagator $K^\sigma$ for an arbitrary  potential $V^\sigma$ is expressed by elementary functions only. 
In the case of potentials \eqref{def:potential-V12-dimensionless} and \eqref{def:potential-V23-dimensionless}
we get
the following propagators
\begin{equation}\label{expr:propagator-V12}
K^{\{1,2\}}(x,y;t)
=\rme^{-2\rmi t}K_{\rm osc}(x,y;t)\left(1-\frac{4\rmi\sin t\left[ xy-\rme^{\rmi t}\right]}{(1+x^2)(1+y^2)} \right),
\end{equation}
\begin{equation}\label{expr:propagator-V23}
K^{\{2,3\}}(x,y;t)
=\rme^{-2\rmi t}K_{\rm osc}(x,y;t)\left(1-\frac{8\rmi\sin t \left[ xy(x^2y^2-3)-3(x^2+y^2)\cos t-3\rmi (x^2y^2+1)\sin t\right]}{(3+x^4)(3+y^4)} \right),
\end{equation}
where
the propagator of the Harmonic oscillator \cite{feynman1948space} is used
\begin{eqnarray}\label{def:oscillator-propagator}
K_{{\rm osc}}(x,y,t)=
\frac{1}{\sqrt{4\pi \rmi \sin t}}\,
\rme^{\frac{\rmi[(x^2+y^2)\cos t-2xy]}{4\sin t}}\,.
\end{eqnarray}

Second, we will define a rational anzatz to compute propagators $K^\sigma$.
In the general case, propagators for the rationally extended oscillators have 
the following structure 
\begin{equation}
K^{\sigma}(x,y;t)=K_{\rm osc}(x,y;t)
\frac{\sum\limits_{k=0}^{\sigma[[-1]]+1}Q_k^{\sigma}(x,y)\rme^{-\rmi kt}}{\sum\limits_{k=0}^{\sigma[[-1]]+1}Q_k^{\sigma}(x,y)}
\end{equation}
where $Q_k^\sigma(x,y)=Q_k^\sigma(y,x)$ are some polynomials 
that can be determined iteratively, which is more efficient than the 
method based on the generating function \cite{pupasov2007exact}. 
The polynomials $Q_k^\sigma(x,y)$ allow also calculate Green functions $G_\sigma(x,y;E)$ 
and generalize Melher's formula for the 
x-Hermite polynomials \cite{gomez2010exceptional}. 

%

\section{Harmonic oscillator, Hermite polynomials and potentials with quasi-equidistant spectrum \label{sesc:Hermonic-oscillator}}

Consider the hamiltonian of the Harmonic oscillator 
\begin{eqnarray}\label{def:Harmonic-oscillator-Hamiltonian-dimensionless-form}
H_{\rm osc}=-\p^2_{xx}+\frac{x^2}{4}\,,
\end{eqnarray}
with eigen-functions 
defined in terms of probabilistic Hermite polynomials
\begin{equation}\label{def:oscillator-eigen-functions}
\psi_n(x)=p_n \mathrm{He}_n(x)\rme^{-\frac{x^2}{4}}\,, \qquad p_n=\left(n!\sqrt{2\pi} \right)^{-\frac{1}{2}}\,.
\end{equation}

Rational extensions are defined as the following perturbations of the Harmonic oscillator \cite{gomez2010exceptional}
\begin{eqnarray}\label{def:rational-extensions-of-Harmonic-oscillator}
H^{\sigma}=-\p^2_{xx}+\frac{x^2}{4}-2\p^2_{xx}(\ln {\rm Wr}[\psi_\sigma(x),x])\,,
\end{eqnarray}
where 
$\sigma$, 
\begin{eqnarray}
\sigma=\{k_1,k_1+1,\ldots,k_{M},k_{M}+1\}\,,\qquad |\sigma|=2M\,,
\end{eqnarray}
is a strictly
increasing sequence of natural numbers, or a Krein-Adle sequence \footnote{We assume that $V_\sigma$ is defined for all 
$x\in \mathbb{R}$, which restricts possible choice of $\sigma$, in particular, $|\sigma|=2M$.
See for the details \cite{gomez2014rational}}.

Following standard notations of the Mathematica program language we denote by $\sigma[[i]]$ and by $\sigma[[-1]]$
 $i$-th element and  the last element of this sequence, respectively.
A sequence of natural numbers appearing as a subscript, for instance, $\psi_{\sigma}(x)$,
implies a set of elements,
\[
\psi_{\sigma}(x)=\psi_{\{n_1,n_2,\ldots,n_{2M}\}}(x)=
\{\psi_{n_1}(x),\psi_{n_2}(x),\ldots,\psi_{n_{2M}}(x)\}\,,
\]
that is, if $(A_n)_{n>0}$ is a sequence of elements, then $A_{\{n_1,\ldots,n_m\}}$ is a set of 
elements.
This agreement allows us to write Wronskians in a compact form
\begin{equation}
{\rm Wr}[\psi_{\sigma}(x),x]=
\left|\begin{array}{cccc}
\psi_{\sigma[[1]]}(x) & \psi_{\sigma[[2]]}(x) & \ldots &\psi_{\sigma[[-1]]}(x)\\
\psi'_{\sigma[[1]]}(x) & \psi'_{\sigma[[2]]}(x) & \ldots &\psi'_{\sigma[[-1]]}(x)\\
\ldots&\ldots&\ldots&\ldots\\
\psi_{\sigma[[1]]}^{(|\sigma|-1)}(x) & \psi_{\sigma[[2]]}^{(|\sigma|-1)}(x) & \ldots &\psi_{\sigma[[-1]]}^{(|\sigma|-1)}(x)\\
\end{array}\right|.
\end{equation}

Hamiltonians $H_{\rm osc}$ and $H^{\sigma}$ can be embedded into a polinomial SUSY algebra 
\cite{andrianov2003nonlinear,andrianov2012nonlinear}
\begin{eqnarray}\label{ident:intertwinning-relation}
LH_{\rm osc} = H^\sigma L\,,\quad L^+L = \prod\limits_{j=1}^{2M}(H_{\rm osc}-\sigma[[j]])\,,\quad LL^+ = \prod\limits_{j=1}^{2M}(H^{\sigma}-\sigma[[j]])\,,
\end{eqnarray}
where $L$ is a differential operator 
 of $2M$-th order \cite{bagrov1995darboux}
\begin{eqnarray}\label{def:intertwinning-operator}
L f(x)=\frac{{\rm Wr}[\psi_{\sigma}(x)\cup \{f(x)\},x]}{{\rm Wr}[\psi_\sigma(x),x]}\,.
\end{eqnarray}

Operator $L$ maps the oscillator eigen-functions \eqref{def:oscillator-eigen-functions}
to the eigen-functions of the rationally extended oscillator 
\begin{equation}\label{def:Rext-oscillator-eigen-functions}
\psi_n^\sigma(x)=N_nL\psi_n(x)\,,
\end{equation}
where a normalization factor is taken into account
\begin{eqnarray}\label{def:normalization-factor}
N_n=\left\lbrace 
\begin{array}{cc}
\left( \prod\limits_{j=1}^{2M}(n-\sigma[[j]])\right) ^{-\frac{1}{2}}\,, & n\notin\sigma\,,\\
0\,, & n\in \sigma\,.
\end{array}
\right.
\end{eqnarray}

Using explicit form of oscillator eigen-functions \eqref{def:oscillator-eigen-functions}
and the following identities 
\[
 {\rm Wr}[\psi_\sigma(x),x]= \rme^{-\frac{Mx^2}{2}}{\rm Wr}[\mathrm{He}_\sigma(x),x]\prod\limits_{n=1}^{2M}p_{\sigma[[n]]}
\]
\[
\ln {\rm Wr}[\psi_\sigma(x),x]= -\frac{Mx^2}{2}+\ln\prod\limits_{n=1}^{2M}p_{\sigma[[n]]}+\ln {\rm Wr}[\mathrm{He}_\sigma(x),x]
\]
we can express rationally extended Harmonic oscillators through the Wronskian of probabilistic Hermite
polynomials only
\begin{eqnarray}\label{def:rational-extensions-of-Harmonic-oscillator-Hermite-form}
H_{\sigma}=-\p^2_{xx}+\frac{x^2}{4}-2\p^2_{xx}(\ln {\rm Wr}[\mathrm{He}_\sigma(x),x])+2M\,.
\end{eqnarray}
Note (see for instance \cite{gomez2014rational}) that $ {\rm Wr}[\mathrm{He}_\sigma(x),x]$ (for the chosen class of $\sigma$ ) is the polynomial of $x^2$ and 
\[
{\rm deg}_{x} {\rm Wr}[\mathrm{He}_\sigma(x),x]=\sum\limits_{n=1}^{2M}(\sigma[[n]]-n+1)\,.
\] 

It is convenient to introduce normalized polynomials $h_n(x)$ 
\begin{equation}\label{def:norm-Hermite-polynom}
h_n(x)=p_n \mathrm{He}_n(x)\,,
\end{equation}
and the corresponding normalized exceptional Hermite polynomials 
\begin{equation}\label{def:norm-xHermite-polynom}
h_n^{\sigma}(x)=N_n {\rm Wr}[h_{\sigma}\cup h_n,x]
\end{equation}

Finally, we define compact notations 
\begin{equation}\label{def:Wronskian-sigma-compact}
W(x)={\rm Wr}[\psi_{\sigma}(x),x]\,,\qquad \hat W(x)={\rm Wr}[\mathrm{He}_{\sigma}(x),x]\,,
\end{equation}
\begin{equation}\label{def:Wronskian-sigma-n-compact}
W_n(x)={\rm Wr}[\psi_{\sigma\setminus \{\sigma[[n]]\}}(x),x]\,,\qquad 
\hat W_n(x)={\rm Wr}[\mathrm{He}_{\sigma\setminus \{\sigma[[n]]\}}(x),x]\,,
\end{equation}
\begin{equation}\label{def:hat-L-compact}
\hat L f=\frac{{\rm Wr}[\mathrm{He}_{\sigma}(x)\cup\{f\},x]}{{\rm Wr}[\mathrm{He}_\sigma(x),x]}\,.
\end{equation}
Note that 
\begin{eqnarray}\label{psi2He-transition}
W(x)&=&\rme^{-\frac{Mx^2}{2}}\hat W(x)\prod\limits_{n=1}^{2M}p_{\sigma[[n]]}\,,\qquad 
W_n(x)=\rme^{-\frac{(2M-1)x^2}{4}}\hat W_n(x)p_{\sigma[[n]]}^{-1}\prod\limits_{j=1}^{2M}p_{\sigma[[j]]}\,,\\
L \rme^{-\frac{x^2}{4}}f &=& \rme^{-\frac{x^2}{4}}\hat L f\,.
\end{eqnarray}

\section{Propagators of 
rationally extended harmonic oscillators \label{sec:Gen-Func-Form}}
\subsection{Generating function formalism}

The Schr\"{o}dinger equation for the Green function reads
\begin{equation}\label{def:Schrodinger-equation-for-Green-function-dimensionless-form}
(\rmi \p_t-H)_xK(x,y;t)=0\,,\qquad K(x,y,0)=\delta(x-y)\,.
\end{equation}

If two Hamiltonians $H_0$ and $H_N$ are related by $N$-th order 
Darboux transformation which remove $N$ levels from the spectrum of 
$H_0$, then corresponding propagators $K_0$ and $K_N$ are related as follows
\cite{pupasov2007exact},
\begin{eqnarray}\label{theorem:transformation-GF-general}
K_N(x,y;t)
     & =& L_{x}
  \sum_{n=1}^{N} (-1)^{n}
  \frac{W_{n}(y)}{W(y)}
  \int_{y}^b K_0(x,z;t)u_n(z)dz\,.
\end{eqnarray}

In the case of rationally extended Harmonic oscillators 
$b=\infty$, $N=|\sigma|$, and the transformation solutions
coincide with Harmonic oscillator eigenfunctions
$u_n=\psi_{\sigma[[n]]}(z)$.

Using \eqref{psi2He-transition} we can replace Wronskians of wave functions by Wronskians of Hermite polynomials 
\begin{eqnarray}\label{transformed-Green-function-isochronous}
K^\sigma(x,y;t)=\sum_{n=1}^{2M}(-1)^{n}
\rme^{\frac{y^2}{4}}\frac{\hat W_{n}(y)}{\hat W(y)}
L_{x}\int_{y}^\infty   K_{\rm osc}(x,z;t)\rme^{-\frac{z^2}{4}}\mathrm{He}_{\sigma[[n]]}(z)dz
\end{eqnarray}

The occurring integrals
$\int_y^{\infty}K_{\rm osc}(x,z,t)\rme^{-\frac{z^2}{4}}\mathrm{He}_n(z)dz$ can be represented as
derivatives of the generating function with respect to the auxiliary current $J$ 
\[
\int_y^{\infty}K_{\rm osc}(x,z,t)\rme^{-\frac{z^2}{4}}\mathrm{He}_{\sigma[[n]]}(z)dz
=\left[\mathrm{He}_{\sigma[[n]]}\left(\p_J\right)S(J)\right]_{J=0}
=\left[\sum\limits_{k=0}^{\sigma[[n]]}h_{\sigma[[n]],k}\frac{\p^k S(J)}{\p J^k}\right]_{J=0}\,,
\]
where $h_{m,k}$ are coefficients of the $\mathrm{He}_m$.
The
generating function reads
\begin{equation}\label{def:generating-function}
S(J|x,y,t)=
\frac{1}{2}
\rme^{\left(\frac{-\rmi t}{2}-\frac{x^2}{4}\right)}
R\left[ \rmi J\sqrt{2 \rmi \sin t}\rme^{-\frac{\rmi t}{2}},\frac{x \rme^{-\frac{\rmi t}{2}}}{\rmi\sqrt{2\rmi\sin t}}\right]E[J,x,y,t]
\end{equation}
where
\[
R\left[ \rmi J\sqrt{2 \rmi \sin t}\rme^{-\frac{\rmi t}{2}},\frac{x \rme^{-\frac{\rmi t}{2}}}{\rmi\sqrt{2\rmi\sin t}}\right] ={\exp}{\left(
J(\rmi J\sin t+x)\exp({-it})\right)}\,,
\]
\[
E[J,x,y,t]={\left(1+\erf\left[
-J\sqrt{\rmi \sin t}\rme^{\frac{-\rmi t}{2}}
-\frac{\rmi\sqrt{\rmi}}{2\sqrt{\sin t}}\left(y\rme^{\frac{\rmi t}{2}}-x\rme^{-\frac{\rmi t}{2}}
\right)\right]\right)}\,,
\]
\[
\erf(z)=\frac{2}{\sqrt{\pi}}\int\limits_{0}^{z}\exp(-t^2)dt\,.
\]
Function $R$ is the generating function of rescaled Hermite polynomials (see also Appendix A)
\[
R[z\sqrt{\alpha},\frac{x}{\sqrt{\alpha}}]={\exp}{\left(xz-\frac{\alpha z^2}{2}\right)}=\sum\limits_{n=0}^{\infty}\mathrm{He}_n^{[\alpha]}(x)\frac{z^n}{n!}\,.
\]
For the compact writing we define 
\begin{equation}\label{def:constants-a-alpha}
\alpha=-2\rmi \sin t\rme^{\rmi t}=1-\rme^{2\rmi t}\,,\qquad 
\rme^{-\frac{\rmi t}{2}}
\sqrt{\rmi\sin t}=\frac{1}{\sqrt{2}}\left( 1-\rme^{-2\rmi t}\right) ^{\frac{1}{2}}\,.
\end{equation}

In what follows we need derivatives of generating functions $R$, $E$
and 
\[
E_0[x,y;t]=E[0,x,y,t]={\left(1+\erf\left[\frac{\rmi\sqrt{\rmi}}{2\sqrt{\sin t}}\left(x\rme^{-\frac{\rmi t}{2}}-y\rme^{\frac{\rmi t}{2}}
\right)
\right]\right)}\,.
\]
{1. \it J-derivatives of R-function}\\
\begin{eqnarray}\label{rel:J-derivative-R}
\frac{k!}{(k-m)!}\left(\p_J^{k-m}
R[J\sqrt{\alpha}\rme^{-\rmi t},\frac{x}{\sqrt{\alpha}}]\right)_{J=0}= 
\frac{k!}{(k-m)!}\mathrm{He}^{[\alpha]}_{k-m}\left(x \right) 
\rme^{-\rmi (k-m) t}
=
\rme^{-\rmi (k-m) t}
\frac{d^m}{dx^m}\mathrm{He}_k^{[\alpha]}\left(x\right) 
\end{eqnarray}
{2. \it J-derivatives of E-function}\\
\begin{eqnarray}\label{rel:J-derivative-E}
\left( \frac{\p^{m+1} E[J]}{\p J^{m+1}}\right)_{J=0}=
K_{\rm osc}(x,y;t)\rme^{\frac{x^2-y^2}{4}}\rme^{\frac{-\rmi t}{2}}2\rmi \sin t \sum\limits_{j=0}^{m}q_{j,m+1}(x,y)\rme^{-\rmi t j}
\end{eqnarray}
{3. \it x-derivatives of $E_0$-function}\\
\begin{eqnarray}\label{rel:x-derivative-E0}
\frac{\p^{k+1} E_0[x,y,t]}{\p x^{k+1}}=
K_{\rm osc}(x,y;t)\rme^{\frac{x^2-y^2}{4}}\rme^{\frac{-\rmi t}{2}}
\left( \frac{1}{2\rmi \sin t}\right)^k 
\sum\limits_{j=0}^{k}w_{j,m+1}(x,y)\rme^{-\rmi t j}
\end{eqnarray}
{4. \it Mixed derivatives of $E$-function}\\
\begin{eqnarray}\label{rel:xJ-derivative-E}
\frac{\p^{k} }{\p x^{k}}\left( \frac{\p^{m+1} E[J]}{\p J^{m+1}}\right)_{J=0} = 
\left( \frac{1}{2\rmi \sin t}\right)^{k-1}
\rme^{\frac{-\rmi t}{2}}
K_{\rm osc}(x,y;t)\rme^{\frac{x^2-y^2}{4}}\sum\limits_{j=0}^{m+k}q_{j,m+k+1}(x,y)\rme^{-\rmi t j}\,.
\end{eqnarray}
In the above expressions $q_{i,j}$ and $w_{k,m}$ are some polynomials.

Now we will substitute these derivatives to calculate
\begin{eqnarray}\label{formula:part-of-sum}
\left[\mathrm{He}_{\sigma[[j]]}\left(\p_J\right)S(J)\right]_{J=0}=
\frac{{\rme}^{\frac{-\rmi t}{2}-\frac{x^2}{4}}}{2}
\left[\sum\limits_{k=0}^{\sigma[[j]]}h_{\sigma[[j]],k}\sum\limits_{m=0}^{k}C_k^m \left(\p_J^{k-m}
R[J\sqrt{\alpha}\rme^{-\rmi t},\frac{x}{\sqrt{\alpha}}]\right) 
\left(\p_J^{m} E[J,x,y,t]\right) 
\right]_{J=0}\,.
\end{eqnarray}

Consider first the following double sum
\[
\left[\sum\limits_{k=0}^{\sigma[[j]]}h_{\sigma[[j]],k}\sum\limits_{m=0}^{k}\frac{k!}{m!(k-m)!} \left(\p_J^{k-m}
R[J\sqrt{\alpha}\rme^{-\rmi t},\frac{x}{\sqrt{\alpha}}]\right) 
\left(\p_J^{m} E[J,x,y,t]\right) 
\right]_{J=0}
\]

We will change the order of summation ($k,m\to m,k$), that is we will fix $m$ and calculate first 
sum by $k\in(\sigma[[j]]-m,\sigma[[j]])$.
\[
\left[\sum\limits_{k=0}^{\sigma[[j]]}\sum\limits_{m=0}^{k} h_{\sigma[[j]],k} \frac{1}{m!}\rme^{-\rmi (k-m) t}
\frac{d^m}{dx^m}\mathrm{He}_k^{[\alpha]}\left(x\right)
\left(\p_J^{m} E[J,x,y,t]\right) 
\right]_{J=0}=
\]
\[
\left[\sum\limits_{m=0}^{\sigma[[j]]} \frac{1}{m!} \left(\p_J^{m} E[J,x,y,t]\right)
\sum\limits_{k=m}^{\sigma[[j]]} h_{\sigma[[j]],k} 
\rme^{-\rmi (k-m) t}\frac{d^m}{dx^m}\mathrm{He}_k^{[\alpha]}\left(x\right)
\right]_{J=0}.
\]
In the last expression we can change the inferior limit $k=m$ to $k=0$ since $\frac{d^m}{dx^m}\mathrm{He}_k^{[\alpha]}\left(x\right)=0$ when $k<m$,
\[
\rme^{-\rmi \sigma[[j]]t}\left[\sum\limits_{m=0}^{\sigma[[j]]} \frac{1}{m!} \left(\p_J^{m} E[J,x,y,t]\right) 
\rme^{\rmi m t} \frac{d^m}{dx^m}\sum\limits_{k=0}^{\sigma[[j]]} h_{\sigma[[j]],k} 
(\rme^{2\rmi t})^{\frac{\sigma[[j]]-k}{2}}\mathrm{He}_k^{[\alpha]}\left(x\right)
\right]_{J=0}.
\]
Afterwards we note that $h_{\sigma[[j]],k} 
(\rme^{2\rmi t})^{\frac{\sigma[[j]]-k}{2}}=h_{\sigma[[j]],k}^{[1-\alpha]}$ (see Appendix A, \eqref{def:rescaled-Hermite-coeffitients}),
\[
\rme^{-\rmi \sigma[[j]]t}\left[\sum\limits_{m=0}^{\sigma[[j]]} \frac{1}{m!} \left(\p_J^{m} E[J,x,y,t]\right) 
\rme^{\rmi m t}
\frac{d^m}{dx^m}\sum\limits_{k=0}^{\sigma[[j]]}h_{\sigma[[j]],k}^{[1-\alpha]}
\mathrm{He}^{[\alpha]}_k\left(x\right)
\right]_{J=0}.
\]
The sum by $k$ represent the umbral composition \eqref{app:umbral-composition} for the generalized Hermite polynomials 
$\mathrm{He}^{[\alpha]}_k\left(x\right)$ \cite{roman1978umbral}, which yields
\[
\rme^{-\rmi \sigma[[j]]t}\left[\sum\limits_{m=0}^{\sigma[[j]]} \frac{1}{m!} \left(\p_J^{m} E[J,x,y,t]\right) 
\rme^{\rmi m t}
\frac{d^m}{dx^m}\mathrm{He}_{\sigma[[j]]}\left(x\right)
\right]_{J=0}\,.
\]

As a result we obtain the following intermediate expression for
the propagator 
\[
K^\sigma=\frac{{\rme}^\frac{-\rmi t}{2}}{2}
\sum_{j=1}^{2M}(-1)^{j}
\rme^{\frac{y^2}{4}}\frac{\hat W_{j}(y)}{\hat W(y)}L_{x}{\rme}^\frac{-x^2}{4}\sum\limits_{m=0}^{\sigma[[j]]} \frac{1}{m!} \left(\p_J^{m} E[J,x,y,t]\right)_{J=0}
\rme^{-\rmi (\sigma[[j]]-m)t} \frac{d^m}{dx^m}\mathrm{He}_{\sigma[[j]]}\left(x\right)\,.
\]
We split the last expression in two terms $K^\sigma=K_E+K_R$, where the first term 
contains error function in $E_0(x,y;t)$ whereas the second term contains elementary functions only
\[
K_E=\frac{{\rme}^\frac{-\rmi t}{2}}{2}
\sum_{j=1}^{2M}(-1)^{j}
\rme^{\frac{y^2}{4}}\frac{\hat W_{j}(y)}{\hat W(y)}L_{x}{\rme}^\frac{-x^2}{4}\rme^{-\rmi \sigma[[j]]t} E_0[x,y,t]
\mathrm{He}_{\sigma[[j]]}\left(x\right)\,,
\]
\[
K_R=\frac{{\rme}^\frac{-\rmi t}{2}}{2}
\sum_{j=1}^{2M}(-1)^{j}
\rme^{\frac{y^2}{4}}\frac{\hat W_{j}(y)}{\hat W(y)}L_{x}{\rme}^\frac{-x^2}{4}\rme^{-\rmi \sigma[[j]]t}\sum\limits_{m=1}^{\sigma[[j]]} \frac{1}{m!} \left(\p_J^{m} E[J,x,y,t]\right)_{J=0}
\rme^{\rmi m t}
\frac{d^m}{dx^m}\mathrm{He}_{\sigma[[j]]}\left(x\right).
\]
Changing operator $L$ by $\hat L$ with the aids of \eqref{psi2He-transition} we get
\[
K_E=\frac{{\rme}^\frac{-\rmi t}{2}}{2}
\sum_{j=1}^{2M}(-1)^{j}
\rme^{\frac{y^2-x^2}{4}}\rme^{-\rmi \sigma[[j]]t}\frac{\hat W_{j}(y)}{\hat W(y)}\hat L_{x} E_0[x,y,t]
\mathrm{He}_{\sigma[[j]]}\left(x\right),
\]
\[
K_R=\frac{{\rme}^\frac{-\rmi t}{2}}{2}
\sum_{j=1}^{2M}(-1)^{j}
\rme^{\frac{y^2-x^2}{4}}\frac{\hat W_{j}(y)}{\hat W(y)}\hat L_{x}\sum\limits_{m=1}^{\sigma[[j]]} \frac{\rme^{-\rmi (\sigma[[j]]-m)t}}{m!} \left(\p_J^{m} E[J,x,y,t]\right)_{J=0}
\frac{d^m}{dx^m}\mathrm{He}_{\sigma[[j]]}\left(x\right).
\]
Let us consider the first term $K_E$.
Though the function $E_0$ contains non-elementary error function $\erf $ it can be seen 
that $K_{E}(x,y;t)$ is expressed in terms of elementary functions only. The error function 
will be cancelled due to the operator $\hat L$ as follows
\[
K_E=\frac{{\rme}^\frac{-\rmi t}{2}}{2}
\sum_{j=1}^{2M}(-1)^{j}
\rme^{\frac{y^2-x^2}{4}}\rme^{-\rmi \sigma[[j]]t}\frac{\hat W_{j}(y)}{\hat W(y)}\hat L_{x} E_0[x,y,t]
\mathrm{He}_{\sigma[[j]]}\left(x\right)=
\]
\[
\frac{{\rme}^\frac{-\rmi t}{2}}{2}
\sum_{j=1}^{2M}(-1)^{j}
\rme^{\frac{y^2-x^2}{4}}\rme^{-\rmi \sigma[[j]]t}\frac{\hat W_{j}(y)}{\hat W(y)} 
\left( E_0[x,y,t] \hat L_{x} 
\mathrm{He}_{\sigma[[j]]}\left(x\right)
+\frac{\sum\limits_{n=1}^{2M}A_{n,j}(x)\p_x^n E_0(x,y;t)}{\hat W(x)}
\right),
\]
where $A_{n,j}(x)$ are some polynomials in $x$.
By the definition \eqref{def:hat-L-compact} of the operator $\hat L$ we have $\hat L_{x} 
\mathrm{He}_{\sigma[[j]]}\left(x\right)=0$, therefore
\[
K_E=\frac{{\rme}^\frac{-\rmi t}{2}\rme^{\frac{y^2-x^2}{4}}}{2\hat W(y)\hat W(x)}
\sum_{j=1}^{2M}(-1)^{j}
\rme^{-\rmi \sigma[[j]]t}\hat W_{j}(y)
\sum\limits_{n=0}^{2M-1}A_{n,j}(x)\p_x^{n+1} E_0(x,y;t)=
\]
\[
\frac{{\rme}^\frac{-\rmi t}{2}\rme^{\frac{y^2-x^2}{4}}}{2\hat W(y)\hat W(x)}
K_{\rm osc}(x,y;t)\rme^{\frac{x^2-y^2}{4}}\rme^{\frac{-\rmi t}{2}}
\sum_{j=1}^{2M}(-1)^{j}
\rme^{-\rmi \sigma[[j]]t}\hat W_{j}(y)
\sum\limits_{n=0}^{2M-1}A_{n,j}(x)
\left( \frac{1}{2\rmi \sin t}\right)^n 
\sum\limits_{l=0}^{n}w_{l,n+1}(x,y)\rme^{-\rmi t l}=
\]
\[
\frac{K_{\rm osc}(x,y;t){\rme}^{-\rmi t}}{2\hat W(y)\hat W(x)}
\sum_{j=1}^{2M}(-1)^{j}
\rme^{-\rmi \sigma[[j]]t}\hat W_{j}(y)
\sum\limits_{n=0}^{2M-1}A_{n,j}(x)
\left( \frac{1}{2\rmi \sin t}\right)^n 
\sum\limits_{l=0}^{n}w_{l,n+1}(x,y)\rme^{-\rmi t l}\,.
\]
Thus we proved that $K_E$ contains elementary functions only. 
Denote $\rme^{-\rmi t}$ by $\lambda$. We can further simplify $K_E$ writing it as 
a product of $K_{\rm osc}$ and a rational function of $x$, $y$ and $\lambda$,
\[
K_E=\frac{K_{\rm osc}(x,y;t)}{\hat W(y)\hat W(x)}
\lambda\sum_{j=1}^{2M}(-1)^{j}
\lambda^{\sigma[[j]]}\hat W_{j}(y)
\sum\limits_{n=0}^{2M-1}A_{n,j}(x)
\left( \frac{\lambda}{1-\lambda^2}\right)^n 
\sum\limits_{l=0}^{n}w_{l,n+1}(x,y)\lambda^{l}=
\]
\[
\frac{K_{\rm osc}(x,y;t)}{\hat W(y)\hat W(x)}
\frac{\sum\limits_{n=1}^{4M+\sigma[[-1]]-1}\bar Q_{n}(x,y)\lambda^{n}}{(1-\lambda^2)^{2M-1}}\,,
\]
where $\bar Q_{n}(x,y)$ are some polynomials.

Following the same line we will consider the second term. It can be seen that 
structures of $K_E$ and $K_R$ coincide, and $K_R$ reads 
\[
K_R=
\frac{K_{\rm osc}(x,y;t)}{\hat W(y)\hat W(x)}
\frac{\sum\limits_{n=0}^{4M+\sigma[[-1]]-1}\tilde Q_{n}(x,y)\lambda^{n}}{(1-\lambda^2)^{2M-1}}
\]

Combining now $K_E$ and $K_R$  we obtain the following expression 
\[
K^\sigma=\frac{K_{\rm osc}(x,y;t)}{\hat W(y)\hat W(x)}\frac{\sum\limits_{n=0}^{\sigma[[-1]]+1+4M-2}\bar Q_n^{\sigma}(x,y)\lambda^n}{\left(1-\lambda^2\right)^{2M-1} }=
\frac{K_{\rm osc}(x,y;t)}{{\rm Wr}[h_{\sigma}(x),x]{\rm Wr}[h_{\sigma}(y),y]}\frac{\sum\limits_{n=0}^{\sigma[[-1]]+1+4M-2}\bar Q_n^{\sigma}(x,y)\lambda^n}{\left(1-\lambda^2\right)^{2M-1} }
\]
where some constant multiplier ${\rm Wr}[h_{\sigma}(x),x]=C\hat W(x)$ is absorbed by redefinition of polynomials $\bar Q$.

Consider the limit $t\to 0$, where $K^\sigma\to \delta(x-y)$ and $K_{\rm osc}\to \delta(x-y)$. From here it follows that
the rational $\lambda$-depending factor has no pole, therefore 
\[
\frac{\sum\limits_{n}^{\sigma[[-1]]+1+4M-2}\bar Q_n^{\sigma}(x,y)\lambda^n}{\left(1-\lambda^2\right)^{2M-1} }=
\frac{\left(1-\lambda^2\right)^{2M-1}\sum\limits_{n=0}^{\sigma[[-1]]+1} Q_n^{\sigma}(x,y)\lambda^n}{\left(1-\lambda^2\right)^{2M-1} }=\sum\limits_{n=0}^{\sigma[[-1]]+1}Q_n^{\sigma}(x,y)\lambda^n
\]
and
\begin{eqnarray}\label{rel:Q-polynomials-sum}
\sum\limits_{k=0}^{\sigma[[-1]]+1}Q_k^\sigma(x,y)={\rm Wr}[h_{\sigma}(x),x]{\rm Wr}[h_{\sigma}(y),y]\,.
\end{eqnarray}

We finally get a rational anzatz for the propagators 
\begin{equation}\label{def:rational-anzatz-for-propagator}
K^{\sigma}(x,y;t)=\frac{K_{osc}(x,y;t)}{{\rm Wr}[h_{\sigma}(x),x]{\rm Wr}[h_{\sigma}(y),y]} \sum\limits_{k=0}^{\sigma[[-1]]+1}Q^\sigma_k(x,y)\rme^{-\rmi kt}
\end{equation}
where $Q^\sigma_k(x,y)=Q^\sigma_k(y,x)$ are some polynomials to be determined.

\subsection{Rational anzatz for the propagators}

Substituting into the rational anzatz  \eqref{def:rational-anzatz-for-propagator}
expansions of propagators in terms of eigen-functions 
\begin{eqnarray}
\label{def:exp-osc-prop}
K_{\rm osc}(x,y;t) &=&\rme^{-\frac{x^2+y^2}{4}}\sum\limits_{n=0}^{\infty}h_n(x)h_n(y)\lambda^n\,,\qquad \lambda=\rme^{-\rmi t}\,,\\
\label{def:exp-rat-osc-prop}
K_{\sigma}(x,y;t) &=&\frac{\rme^{-\frac{x^2+y^2}{4}}}{{\rm Wr}[h_{\sigma}(x),x]{\rm Wr}[h_{\sigma}(y),y]}
\sum\limits_{n\in \mathbb{N}\backslash {\sigma}}h^{\sigma}_n(x)h^{\sigma}_n(y)\lambda^n
\end{eqnarray}
we obtain a system of equations for the polynomial coefficients 
$Q^\sigma_k(x,y)$.
The solution of this system is given 
by a recursive procedure 
\begin{eqnarray}\label{def:Q-polynomials}
Q^\sigma_k(x,y)=\frac{1}{h_0(x)h_0(y)}\left(h^{\sigma}_k(x)h^{\sigma}_k(y)-\sum\limits_{j=1}^{k}Q^\sigma_{k-j}(x,y)h_{j}(x)h_{j}(y)\right),\qquad 0\leq k\leq \sigma[[-1]]+1\,.
\end{eqnarray}

\subsubsection{Nonlinear connection Lemma for the exceptional Hermite polynomials}
{\it Lemma.} Given a Krein-Adler sequence $\sigma=\{k_1,k_1+1\ldots, k_{M},k_M+1\}$, the corresponding family 
of (formally normalized) polynomials  $h_n^{\sigma}(x)$ obeys the following relation
\begin{eqnarray}\label{def:nonlinear-connection}
\sum\limits_{k=0}^{\sigma[[-1]]+1}h_{m-k}(x)h_{m-k}(y)Q^{\sigma}_k(x,y)=h^\sigma_m(x)h^\sigma_m(y)\,,
\end{eqnarray}
where polynomials $Q^\sigma_k$ are given by \eqref{def:Q-polynomials}.

{\it Proof.} The proof follows 
from \eqref{def:rational-anzatz-for-propagator}, \eqref{def:exp-osc-prop}, \eqref{def:exp-rat-osc-prop} and \eqref{def:Q-polynomials} $\Box$

Relation \eqref{def:nonlinear-connection} can also be written in terms of wave functions
\begin{eqnarray}\label{def:wave-function-connection}
\frac{
\sum\limits_{k=0}^{\sigma[[-1]]+1}\psi_{m-k}(x)\psi_{m-k}(y)Q^{\sigma}_k(x,y)
}{
\sum\limits_{k=0}^{\sigma[[-1]]+1}Q^{\sigma}_k(x,y)}=\psi^\sigma_m(x)\psi^\sigma_m(y)\,.
\end{eqnarray}
 
There are following properties of the $Q^\sigma$-polynomials:\\
1) Symmetry
\[
Q_k^\sigma(x,y)=Q_k^\sigma(y,x)
\]
2)Parity
\[
Q_k^\sigma(-x,-y)=Q_k^\sigma(x,y)
\]
\[
Q^{\{\sigma\}}_{2k}(-x,y)=(-1)^{\left( {\rm deg}h_0^\sigma\right) }Q_{2k}(x,y)\,,\qquad  Q_{2k+1}(-x,y)=-(-1)^{\left( {\rm deg}h_0^\sigma\right) }Q_{2k+1}(x,y)
\]
where ${\rm deg}h_0^\sigma=\sum\limits_{j=1}^{|\sigma|}(\sigma[[j]]-j+1)-|\sigma|$.

In the Appendix B we present an example of a non-Krein-Adler sequence $\sigma=\{1\}$
when the nonlinear connection lemma is also holds. We suppose that \eqref{def:nonlinear-connection}
holds for an arbitrary $\sigma$. 

\subsubsection{x-Mehler formula}
We first recall the Mehler formula 
\begin{eqnarray}\label{rel:Mehler-formula}
\sum\limits_{n=0}^{\infty}\mathrm{He}_n(x)\mathrm{He}_n(y)\frac{\lambda^n}{n!}=\frac{1}{\sqrt{1-\lambda^2}}
\rme^{\frac{-\lambda^2(x^2+y^2)+2\lambda xy}{2(1-\lambda^2)}}\,.
\end{eqnarray}
Using the nonlinear connection lemma \eqref{def:nonlinear-connection} we 
obtain the following generalization of the Mehler formula to the case of 
exceptional Hermite polynomials
\begin{eqnarray}\label{rel:x-Mehler-formula}
\sum\limits_{n=0}^{\infty}h_n^\sigma(x)h_n^\sigma(y)\lambda^n=\frac{1}{\sqrt{2\pi(1-\lambda^2)}}
\rme^{\frac{-\lambda^2(x^2+y^2)+2\lambda xy}{2(1-\lambda^2)}}\sum\limits_{j=0}^{\sigma[[-1]]+1}Q_j^\sigma(x,y)\lambda^j\,.
\end{eqnarray}

\subsubsection{Alternative form of the rationally extended Harmonic oscillators}
Consider integral kernels of Hamiltonian operators
\begin{equation}\label{def:hamiltonian-kernel}
H_{osc}(x,y)=\left[ -\p^2_{xx}+V_{osc}(x)\right]\delta(x-y)=\sum\limits_{n=0}^{\infty}n\psi_n(x)\psi_n(y)\,,
\end{equation}
\begin{equation}\label{def:hamiltonian-kernel}
H^{\sigma}(x,y)=\left[ -\p^2_{xx}+V^{\sigma}(x)\right]\delta(x-y)=\sum\limits_{n=0}^{\infty}n\psi^{\sigma}_n(x)\psi^{\sigma}_n(y)\,,
\end{equation}
using \eqref{def:wave-function-connection} we can represent the second kernel as follows
\begin{equation}\label{def:hamiltonian-kernel}
H^{\sigma}(x,y)=\left[ -\p^2_{xx}+V_{\rm osc}(x)-\frac{\sum\limits_{k=0}^{\sigma[[-1]]+1}k Q_k^\sigma(x,y)}{\sum\limits_{j=0}^{\sigma[[-1]]+1} Q_j^\sigma(x,y)}\right]\delta(x-y)\,.
\end{equation}
From here is follows that
\[
\Delta V^\sigma(x)=-2 \p_{xx}\log\left[ {\rm Wr}[\psi_\sigma(x),x]\right] =-\frac{\sum\limits_{k=0}^{\sigma[[-1]]+1}k Q_k^\sigma(x,x)}{\sum\limits_{j=0}^{\sigma[[-1]]+1} Q_j^\sigma(x,x)}\]

\subsubsection{Green functions}

Consider the Green function (resolvent kernel of the Hamiltonian operator) 
\[
G(x,y;E)=\rmi \int\limits_{0}^{\infty}K(x,y;t)\rme^{\rmi E t}dt
\]
The transformation formula for the propagators implies also 
the following relation for the Green functions
\begin{equation}\label{def:rational-anzatz-for-green-function}
G_{\sigma}(x,y;E)=\frac{1}{W[h_{\sigma}(x),x]W[h_{\sigma}(y),y]}
 \sum\limits_{k=0}^{\sigma[[-1]]+1}Q_k(x,y)G_{\rm osc}(x,y;E-k)
\end{equation}

\subsection{Examples}

Using first and second excited states of Harmonic oscillator
\[
\psi_1(x)=p_1x \rme^{-x^2/4}\qquad \psi_2(x)=p_2(x^2-1)\rme^{-x^2/4}
\]
we
obtain a perturbed Harmonic oscillator potential \cite{carinena2008quantum}
\begin{equation}\label{def:potential-V12-dimensionless-example}
V_{\{1,2\}}[x]=\frac{x^2}{4}+2\left(1 + 2\frac{(x^2-1)}{(x^2+1)^2}\right) 
\end{equation}

Connection polynomials read
\[
Q^{\{1,2\}}_{\{0,1,2,3\}}=
\left\lbrace 
\frac{1}{2\pi}, -\frac{xy}{2\pi}, \frac{x^2y^2+x^2+y^2-1}{4\pi},\frac{xy}{2\pi}\right\rbrace \,.
\]

The propagator for the \Sh equation with Hamiltonian
$K^{(1,2)}=-\p_x^2+V^{\{1,2\}}(x)$ has the following compact expression
\begin{equation}\label{expr:propagator-V12-example}
K^{\{1,2\}}(x,y;t)
=\rme^{-2\rmi t}K_{\rm osc}(x,y;t)\left(1-\frac{4\rmi\sin t\left[ xy-\rme^{\rmi t}\right]}{(1+x^2)(1+y^2)} \right),
\end{equation}

Next simple expression for the propagator we can obtain using second and third excited states of Harmonic oscillator
\[
\psi_2(x)=p_2(x^2-1)\rme^{-x^2/4}\qquad \psi_3(x)=p_3x(x^2-3)\rme^{-x^2/4}\,.
\]
In this case we obtain two-well perturbed Harmonic oscillator potential 
(potentials $V_{\{k,k+1\}}$ have $k$ shallow
minima at their bottom)
\begin{equation}\label{def:potential-V23-dimensionless-example}
V_{\{2,3\}}[x]=\frac{x^2}{4}+2\left(1 +4x^2\frac{x^4-
9}{(x^4+3)^2}\right) 
\end{equation}
Connection polynomials read
\[
Q^{\{2,3\}}_{\{0,1,2,3,4\}}=
\left\lbrace 
\frac{(x^2+1)(y^2+1)}{4\pi},\frac{xy(3-x^2y^2)}{6\pi},\frac{x^4y^4+3x^4+3y^4-12x^2y^2-3}{24\pi},\frac{xy(x^2y^2-3)}{6\pi},
\frac{(x^2-1)(y^2-1)}{4\pi}\right\rbrace \,.
\]

The propagator for the \Sh equation with Hamiltonian
$H^{\{2,3\}}=-\p_x^2+V^{\{2,3\}}(x)$ reads
\begin{equation}\label{expr:propagator-V23-example}
K^{\{2,3\}}(x,y;t)
=\rme^{-2\rmi t}K_{\rm osc}(x,y;t)\left(1-\frac{8\rmi\sin t \left[ xy(x^2y^2-3)-3(x^2+y^2)\cos t-3\rmi (x^2y^2+1)\sin t\right]}{(3+x^4)(3+y^4)} \right),
\end{equation}

\section{Conclusions}
Propogators 
\begin{equation}
K^{\sigma}(x,y;t)=K_{\rm osc}(x,y;t)
\frac{\sum\limits_{k=0}^{\sigma[[-1]]+1}Q_k^{\sigma}(x,y)\rme^{-\rmi kt}}{\sum\limits_{k=0}^{\sigma[[-1]]+1}Q_k^{\sigma}(x,y)}
\end{equation}
present a new example of Feynman path integrals that can be calculated analytically \cite{grosche1998handbook}. 
The key formula \eqref{def:Q-polynomials}
\begin{eqnarray}
Q^\sigma_k(x,y)=\frac{1}{h_0(x)h_0(y)}\left(h^{\sigma}_k(x)h^{\sigma}_k(y)-\sum\limits_{j=1}^{k}Q^\sigma_{k-j}(x,y)h_{j}(x)h_{j}(y)\right),\qquad 0\leq k\leq \sigma[[-1]]+1\,,
\end{eqnarray}
which define nonlinear connection between x-Hermite and Hermite polynomials can be easily realized in any computer algebra system.
One can use these propagators to test various approximations for the corresponding
path integrals. Using these propagators wave-packet dynamics in multi-well potentials can be studied analytically.

\section*{Appendix A. Rescaled Hermite polynomials, Appell sequences and umbral composition \cite{roman1978umbral}}
The generating function of the probabilistic  Hermite polynomials reads
\begin{equation}\label{def:generating-function-Hermite}
R[z,x]={\exp}{\left(x z-\frac{z^2}{2}\right)}=\sum\limits_{n=0}^{\infty}\mathrm{He}_n(x)\frac{z^n}{n!}\,.
\end{equation}
\[
\mathrm{He}_n(x)=\sum\limits_{k=0}^{n}h_{n,k}x^k\,,
\]

Rescaled Hermite polynomials 
\begin{eqnarray}\label{def:rescaled-Hermite-polynomials}
\mathrm{He}_n^{[\alpha]}(x) &=& \alpha^{\frac{n}{2}}\mathrm{He}_n\left(\frac{x}{\sqrt{\alpha}}\right)\,,\\
h_{n,k}^{[\alpha]} &=& \alpha^{\frac{(n-k)}{2}}h_{n,k}\,,\label{def:rescaled-Hermite-coeffitients}
\end{eqnarray}
with the following generating function
\[
R[z\sqrt{\alpha},\frac{x}{\sqrt{\alpha}}]={\exp}{\left(xz-\frac{\alpha z^2}{2}\right)}=\sum\limits_{n=0}^{\infty}\mathrm{He}_n^{[\alpha]}(x)\frac{z^n}{n!}\,.
\]
form an Appell sequence of polynomials.

Let $A_n(x)$ and $B_n(x)$ be two Appel sequences of polynomials \cite{roman1982theory} generated by functions
$S[x,g]$, $R[x,g]$, 
\[
S[x,g]=s(g)\rme^{xg}=\sum\limits_{n=0}^{\infty}A_n(x)\frac{g^n}{n!}\,,
\]
\[
R[x,g]=r(g)\rme^{xg}=\sum\limits_{n=0}^{\infty}B_n(x)\frac{g^n}{n!}\,,
\]
where $A_n(x)=\sum\limits_{k=0}^{n}a_{n,k}x^k$ ,$B_n(x)=\sum\limits_{k=0}^{n}b_{n,k}x^k$.

Define the umbral composition by the following formula 
\begin{equation}
(A_n\circ B)(x)=\sum\limits_{k=0}^{n}a_{n,k}B_k(x)=\sum\limits_{k=0}^{n}\sum\limits_{j=0}^{k}a_{n,k}b_{k,j}x^k\,.
\end{equation}
Then the sequence $C_n(x)=(A_n\circ B)(x)$ also is an Appell sequence. 
Its generating function, denoted by $U[x,g]=S[x,g]\circ R[x,g]$, has the following form
\[
U[x,g]=s(g)r(g)\rme^{xg}=\sum\limits_{n=0}^{\infty}C_n(x)\frac{g^n}{n!}\,.
\]
We can apply these facts to the sequences of generalized Hermite polynomials $He_n^{[\alpha]}(x)$
\[
{\exp}{\left(xz-\frac{\alpha z^2}{2}\right)}\circ{\exp}{\left(xz-\frac{\beta z^2}{2}\right)}=
{\exp}{\left(xz-\frac{(\alpha+\beta)z^2}{2}\right)}\,.
\]
As a result we get the umbral composition of rescaled Hermite polynomials
\begin{equation}\label{app:umbral-composition}
\left( He_n^{[\alpha]}\circ He^{[\beta]}\right) (x)=\sum\limits_{k=0}^{n} h^{[\alpha]}_{n,k}He^{[\beta]}_k(x)=
He_n^{[\alpha+\beta]}(x)\,.
\end{equation}

\section*{Appendix B. $Q$-polynomials for non-Krein-Adler sequence $\sigma$}

Here we consider an example of the polynomial connection 
when $\sigma=\{1\}$, which is a non-Krein-Adler sequence \cite{gomez2014rational}, and hence, when polynomials 
$h_n^{1}$ do not represent a sequence of orthogonal polynomials. 
Nevertheless, by direct calculations we can verify 
that the Non-linear connection lemma is valid in this case.
Consider formally normalized product of two wronskians
\begin{eqnarray}
h^{\{1\}}_m(x)h^{\{1\}}_m(y)=\frac{1}{\sqrt{2\pi}(1-m)}
\left|
\begin{array}{cc}
x & h_m(x)\\
1 & h'_m(x)
\end{array}
\right|
\left|
\begin{array}{cc}
y & h_m(y)\\
1 & h'_m(y)
\end{array}
\right|
\end{eqnarray}
\[
=\frac{1}{\sqrt{2\pi}(1-m)}\left(x\sqrt{m}h_{m-1}(x)-h_m(x)\right) \left(y\sqrt{m}h_{m-1}(y)-h_m(y)\right).
\]
Using the recurrence relation for the normalized probabilistic Hermite polynomials   
\[
x\sqrt{m}h_{m-1}(x)=mh_m(x)+{\sqrt{m(m-1)}} h_{m-2}(x)
\]
we get
\[
h^{\{1\}}_m(x)h^{\{1\}}_m(y)=\frac{1}{\sqrt{2\pi}(1-m)}\left([m-1]h_m(x)+{\sqrt{m(m-1)}} h_{m-2}(x)\right)\left([m-1]h_m(y)+
{\sqrt{m(m-1)}} h_{m-2}(y)\right)
\]
\[
=-\frac{1}{\sqrt{2\pi}}\left(h_m(x)+\frac{\sqrt{m}}{\sqrt{(m-1)}} h_{m-2}(x)\right)\left([m-1]h_m(y)+{\sqrt{m(m-1)}} h_{m-2}(y)\right)
\]
\[
=\frac{-1}{\sqrt{2\pi}}\left(-h_m(x)h_m(y)
+mh_m(x)h_m(y)+
{\sqrt{m(m-1)}}(h_m(x)h_{m-2}(y)+h_m(y)h_{m-2}(x))
+mh_{m-2}(x)h_{m-2}(y)
\right) 
\]
\[
=\frac{1}{\sqrt{2\pi}}h_m(x)h_m(y)
-\frac{1}{\sqrt{2\pi}}(xh_{m-1}(x)-{\sqrt{(m-1)}}h_{m-2}(x))(yh_{m-1}(y)-{\sqrt{(m-1)}}h_{m-2}(y))\]
\[-\frac{1}{\sqrt{2\pi}}{\sqrt{(m-1)}}\left( (xh_{m-1}(x)-{\sqrt{(m-1)}}h_{m-2}(x))h_{m-2}(y)+(yh_{m-1}(y)-{\sqrt{(m-1)}}h_{m-2}(y))h_{m-2}(x)\right) \]\[
-\frac{1}{\sqrt{2\pi}}mh_{m-2}(x)h_{m-2}(y)
\]
\[
=\frac{1}{\sqrt{2\pi}}h_m(x)h_m(y)
-\frac{1}{\sqrt{2\pi}}(xh_{m-1}(x)yh_{m-1}(y)
+{(m-1)}h_{m-2}(x)h_{m-2}(y))\]
\[+\frac{1}{\sqrt{2\pi}}2(m-1) h_{m-2}(x)h_{m-2}(y) \]\[
-\frac{1}{\sqrt{2\pi}}mh_{m-2}(x)h_{m-2}(y)
\]
\[
=\frac{1}{\sqrt{2\pi}}\left(h_m(x)h_m(y)
-xyh_{m-1}(x)h_{m-1}(y)
-h_{m-2}(x)h_{m-2}(y)
\right) 
\]
That is, polynomials $h^{\{1\}}_m(x)h^{\{1\}}_m(y)$ satisfy to the following 3-term representation 
\[
h^{\{1\}}_m(x)h^{\{1\}}_m(y)=\sum\limits_{k=0}^{2}h_{m-k}(x)h_{m-k}(y)Q^{\{1\}}_k(x,y)\,,
\]
where
\[
Q_0^{\{1\}}(x,y)=\frac{1}{\sqrt{2\pi}}\,,
\]
\[
Q_1^{\{1\}}(x,y)=-\frac{xy}{\sqrt{2\pi}}\,,
\]
\[
Q_2^{\{1\}}(x,y)=-\frac{1}{\sqrt{2\pi}}\,.
\]
We also verified by computer algebra that the nonlinear connection Lemma is valid for arbitrary $\sigma$ with 
$\sigma[[-1]]<5$.


\begin{thebibliography}{10}

\bibitem{gomez2014rational}
D. G{\'o}mez-Ullate, Y. Grandati, and R. Milson.
\newblock Rational extensions of the quantum harmonic oscillator and
  exceptional hermite polynomials.
\newblock {\em Journal of Physics A: Mathematical and Theoretical},
  47(1):015203, 2014.

\bibitem{carinena2007isochronous}
J.F.~Carinena, A.~M. Perelomov, and M.F.~Ranada.
\newblock Isochronous classical systems and quantum systems with equally spaced
  spectra.
\newblock In {\em Journal of Physics: Conference Series}, volume~87, page
  012007. IOP Publishing, 2007.

\bibitem{dubov1994equidistant}
S.~Yu. Dubov, V.M.~Eleonskii, and N.E.~Kulagin.
\newblock Equidistant spectra of anharmonic oscillators.
\newblock {\em Chaos: An Interdisciplinary Journal of Nonlinear Science},
  4(1):47--53, 1994.

\bibitem{bagrov1997darboux}
V.G.~Bagrov and B.F.~Samsonov.
\newblock Darboux transformation of the schr{\"o}dinger equation.
\newblock {\em Physics of Particles and Nuclei}, 28(4):374--397, 1997.

\bibitem{andrianov2012nonlinear}
A.A.~Andrianov and M.V.~Ioffe.
\newblock Nonlinear supersymmetric quantum mechanics: concepts and
  realizations.
\newblock {\em Journal of Physics A: Mathematical and Theoretical},
  45(50):503001, 2012.

\bibitem{das1990propagators}
A. Das and W.-J. Huang.
\newblock Propagators for shape-invariant potentials.
\newblock {\em Phys. Rev. D}, 41:3241--3247, May 1990.

\bibitem{PhysRevD.47.4796}
A.~M. Jayannavar and A. Khare.
\newblock Propagators for shape-invariant singular potentials.
\newblock {\em Phys. Rev. D}, 47:4796--4797, May 1993.

\bibitem{pupasov2005exact}
A.~M. Pupasov, B.F. Samsonov .
\newblock Exact propagators for soliton potentials.
\newblock {\em SIGMA. Symmetry, Integrability and Geometry: Methods and
  Applications}, 1:020, 2005.

\bibitem{samsonov2005susy}
B.F.~Samsonov, C.V.~Sukumar, and A.M.~Pupasov.
\newblock SUSY transformation of the green function and a trace formula.
\newblock {\em Journal of Physics A: Mathematical and General}, 38(34):7557,
  2005.

\bibitem{samsonov2006exact}
B.F. Samsonov and A.M. Pupasov.
\newblock Exact propagators for complex SUSY partners of real potentials.
\newblock {\em Physics Letters A}, 356(3):210--214, 2006.

\bibitem{pupasov2007exact}
A.M. Pupasov, B.F. Samsonov, and U. G{\"u}nther.
\newblock Exact propagators for SUSY partners.
\newblock {\em Journal of physics A: Mathematical and Theoretical},
  40(34):10557, 2007.

\bibitem{feynman1948space}
R.~P.\ Feynman.
\newblock Space-time approach to non-relativistic quantum mechanics.
\newblock {\em Reviews of Modern Physics}, 20(2):367, 1948.

\bibitem{gomez2010exceptional}
D. Gomez-Ullate, N. Kamran, and R. Milson.
\newblock Exceptional orthogonal polynomials and the Darboux transformation.
\newblock {\em Journal of Physics A: Mathematical and Theoretical},
  43(43):434016, 2010.

\bibitem{andrianov2003nonlinear}
A.A.~Andrianov and A.V.~Sokolov.
\newblock Nonlinear supersymmetry in quantum mechanics: algebraic properties
  and differential representation.
\newblock {\em Nuclear Physics B}, 660(1):25--50, 2003.

\bibitem{bagrov1995darboux}
V.G. Bagrov and B.F. Samsonov.
\newblock Darboux transformation, factorization, and supersymmetry in
  one-dimensional quantum mechanics.
\newblock {\em Theoretical and Mathematical Physics}, 104(2):1051--1060, 1995.

\bibitem{roman1978umbral}
S. Roman.
\newblock
The umbral calculus.
\newblock {\em Advances in Mathematics}, 27(2):95 -- 188, 1978.

\bibitem{carinena2008quantum}
J.F.~Carinena, A.M.~Perelomov, M.F.~Ranada, and M.~Santander.
\newblock A quantum exactly solvable nonlinear oscillator related to the
  isotonic oscillator.
\newblock {\em Journal of Physics A: Mathematical and Theoretical},
  41(8):085301, 2008.

\bibitem{grosche1998handbook}
C. Grosche and F. Steiner.
\newblock Handbook of Feynman path integrals.
\newblock {\em Springer tracts in modern physics}, 145:1, 1998.

\bibitem{roman1982theory}
S. Roman.
\newblock The theory of the umbral calculus. I.
\newblock {\em Journal of mathematical analysis and applications},
  87(1):58--115, 1982.

\end{thebibliography}
\end{document}